\documentclass[journal=jpclcd,manuscript=letter]{achemso}
\usepackage[version=3]{mhchem} 
\usepackage[T1]{fontenc}       
\usepackage{color}
\usepackage{pdfpages}
\author{Christian Vorwerk}
\email{vorwerk@physik.hu-berlin.de}
\affiliation[Humboldt-Universit\"at zu Berlin]
{Institut f\"ur Physik and IRIS Adlershof, Humboldt-Universit\"at zu Berlin, 12489 Berlin, Germany}
\alsoaffiliation{European Theoretical Spectroscopy Facility}
\author{Claudia Hartmann}
\affiliation{Renewable Energy, Helmholtz-Zentrum Berlin f\"ur Materialien und Energie GmbH, 14109 Berlin, Germany}
\author{Caterina Cocchi}
\affiliation[Humboldt-Universit\"at zu Berlin]
{Institut f\"ur Physik and IRIS Adlershof, Humboldt-Universit\"at zu Berlin, 12489 Berlin, Germany}
\alsoaffiliation{European Theoretical Spectroscopy Facility}
\author{Golnaz Sadoughi}
\affiliation{Clarendon Laboratory, Department of Physics, University of Oxford, Oxford OX1 3PU, United Kingdom}
\author{Severin N. Habisreutinger}
\affiliation{Clarendon Laboratory, Department of Physics, University of Oxford, Oxford OX1 3PU, United Kingdom}
\altaffiliation{Chemistry and Nanoscience Center, National Renewable Energy Laboratory (NREL), Golden, CO, USA}
\author{Roberto F\'{e}lix}
\affiliation{Renewable Energy, Helmholtz-Zentrum Berlin f\"ur Materialien und Energie GmbH, 14109 Berlin, Germany}
\author{Regan G. Wilks}
\affiliation{Renewable Energy, Helmholtz-Zentrum Berlin f\"ur Materialien und Energie GmbH, 14109 Berlin, Germany}
\alsoaffiliation{Energy Materials In-Situ Laboratory Berlin (EMIL), Helmholtz-Zentrum Berlin f\"ur Materialien und Energie GmbH, 12489 Berlin, Germany}
\author{Henry J. Snaith}
\affiliation{Clarendon Laboratory, Department of Physics, University of Oxford, Oxford OX1 3PU, United Kingdom}
\author{Marcus B\"ar}
\affiliation{Renewable Energy, Helmholtz-Zentrum Berlin f\"ur Materialien und Energie GmbH, 14109 Berlin, Germany}
\alsoaffiliation{Energy Materials In-Situ Laboratory Berlin (EMIL), Helmholtz-Zentrum Berlin f\"ur Materialien und Energie GmbH, 12489 Berlin, Germany}
\author{Claudia Draxl}
\affiliation[Humboldt-Universit\"at zu Berlin]
{Institut f\"ur Physik and IRIS Adlershof, Humboldt-Universit\"at zu Berlin, 12489 Berlin, Germany}
\alsoaffiliation{European Theoretical Spectroscopy Facility}

\title
  {Exciton-Dominated Core-Level Absorption Spectra of Hybrid Organic-Inorganic Lead Halide Perovskites}

\abbreviations{IR,NMR,UV}

\begin{document}

\begin{tocentry}





\includegraphics[width=1.0\linewidth]{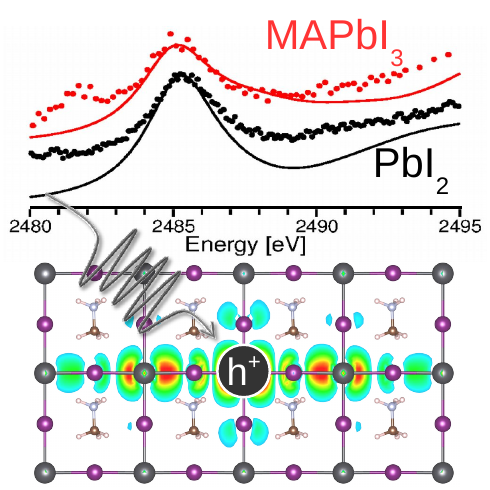}

\end{tocentry}

\begin{abstract}
In a combined theoretical and experimental work, we investigate X-ray Absorption Near-Edge Structure (XANES) spectroscopy of the I $L_3$ and the Pb $M_5$ edges of the methylammonium lead iodide (\ce{MAPbI3}) hybrid inorganic-organic perovskite and its binary phase \ce{PbI2}. The absorption onsets are dominated by bound excitons with sizable binding energies of a few hundred meV and pronounced anisotropy.
The spectra of both materials exhibit remarkable similarities, suggesting that the fingerprints of core excitations in \ce{MAPbI3} are essentially given by its inorganic component, with negligible influence from the organic groups. The theoretical analysis complementing experimental observations provides the conceptual insights required for a full characterization of this complex material.
\end{abstract}

Hybrid organic-inorganic halide perovskites have emerged in the last few years as photovoltaic materials with remarkable efficiencies. Starting from the seminal work by Kojima \textit{et al.} in 2009~\cite{MAPbI3seminal}, solar cells based on hybrid perovskites have reached power conversion efficiencies (PCE) close to that of silicon wafer-based devices~\cite{solar,solar2,solar3,yin+15jmca}, making them the \textit{"next big thing in photovoltaics"} \cite{bigthing}. 
This astonishingly fast development has sparked great interest in the structural, electronic, and optical properties of the hybrid organic-inorganic halide perovskites \cite{Stoumpous, exp_excitonbind,bren+16natrm, Schulz2014}, also stimulating the theoretical community \cite{MAPbI3optic, Zu2014,MAPbI3opticalGW,MAPbI3opticalGW2,MAPbI3opticalDFT,Menendez,li-rink16prb,gao+16prb, Kresse2016,Giustino2015,Giustino2017, MAPbI3GW2}.
While the PCE achieved in laboratory environments is encouraging in view of commercial applications, the long-term stability of the samples remains a critical issue \cite{BertrandeChemMater2015, YangInorgChem2017, LeijtensAdvEnergyMat2015, TiepAdvEnergyMat2016}. 
For methylammonium lead iodide (\ce{MAPbI3}), one of the most studied metal-halide perovskites, previous X-ray photoelectron spectroscopy studies have identified sample degradation due to light irridation, both in the visible \cite{degradation:light} and in the X-ray \cite{degradation:x-ray, bestpractices} region, as well as due to exposure to air and moisture \cite{degradation:water}.
Under these conditions, \ce{MAPbI3} may decompose into the binary phase \ce{PbI2}. Additional decomposition to elemental metallic Pb and \ce{I2} under X-ray illumination was recently reported~\cite{degradation:light, YangInorgChem2017}. 

Access to the local electronic and chemical structure can be achieved with X-ray Absorption Near-Edge Structure (XANES) spectroscopy. 
In this technique, the absorption of X-ray radiation in resonance with an atomic absorption edge yields information about the local environment surrounding the absorbing atom. 
Although XANES provides relevant species-specific information about phase separation and the electronic structure of the sample, the interpretation and rationalization of the spectra requires additional insight. In particular, an open question is whether the presence of degradation products, such as the binary phase \ce{PbI2}, can be determined by X-ray absorption spectroscopy. The required analysis is provided by \textit{ab initio} many-body theory, which combines an accurate description of the electronic structure of the system with an explicit treatment of excitonic effects, that can be crucial in the absorption of X-ray radiation from core electrons. Theoretical studies based on this methodology have been performed to determine the response of \ce{MAPbI3} to visible light \cite{MAPbI3optic,MAPbI3GW2, Kresse2016,Zu2014}. In optical spectroscopy, transitions from valence to conduction bands are probed, and valence excitons may form due to the Coloumb interaction between these states. Core level spectroscopy provides complementary information. With the selective excitation of the initial deep-lying states, core level spectroscopy allows a direct probing of the local environment of the excited species~\cite{Weine2009,Cocchi2016,Fossard2016,Schwartz2016,Cocchi2015}. First-principle theory, in turn, enables a thorough characterization of the excitations in terms of band contributions and spatial extension.

In this Letter, we present the results of a joint theoretical and experimental work, where we investigate XANES and core excitations from the I $L_{3}$ and the Pb $M_{5}$ edges in \ce{MAPbI3} and \ce{PbI2}. 
By exploring differences and similarities between the XANES of the perovskite and of its binary phase \ce{PbI2}, we discuss the role of the inorganic part of the hybrid compound in determining the absorption behavior of \ce{MAPbI3}.
We perform a careful analysis of the measured and computed spectra, focusing on the main features at the absorption onset, where excitonic effects are especially pronounced.
A detailed inspection of the character of the lowest-energy electron-hole pairs reveals their strong anisotropy and their sizable binding energy of a few hundreds meV.
\begin{figure}[t]
\includegraphics[width=0.425\linewidth]{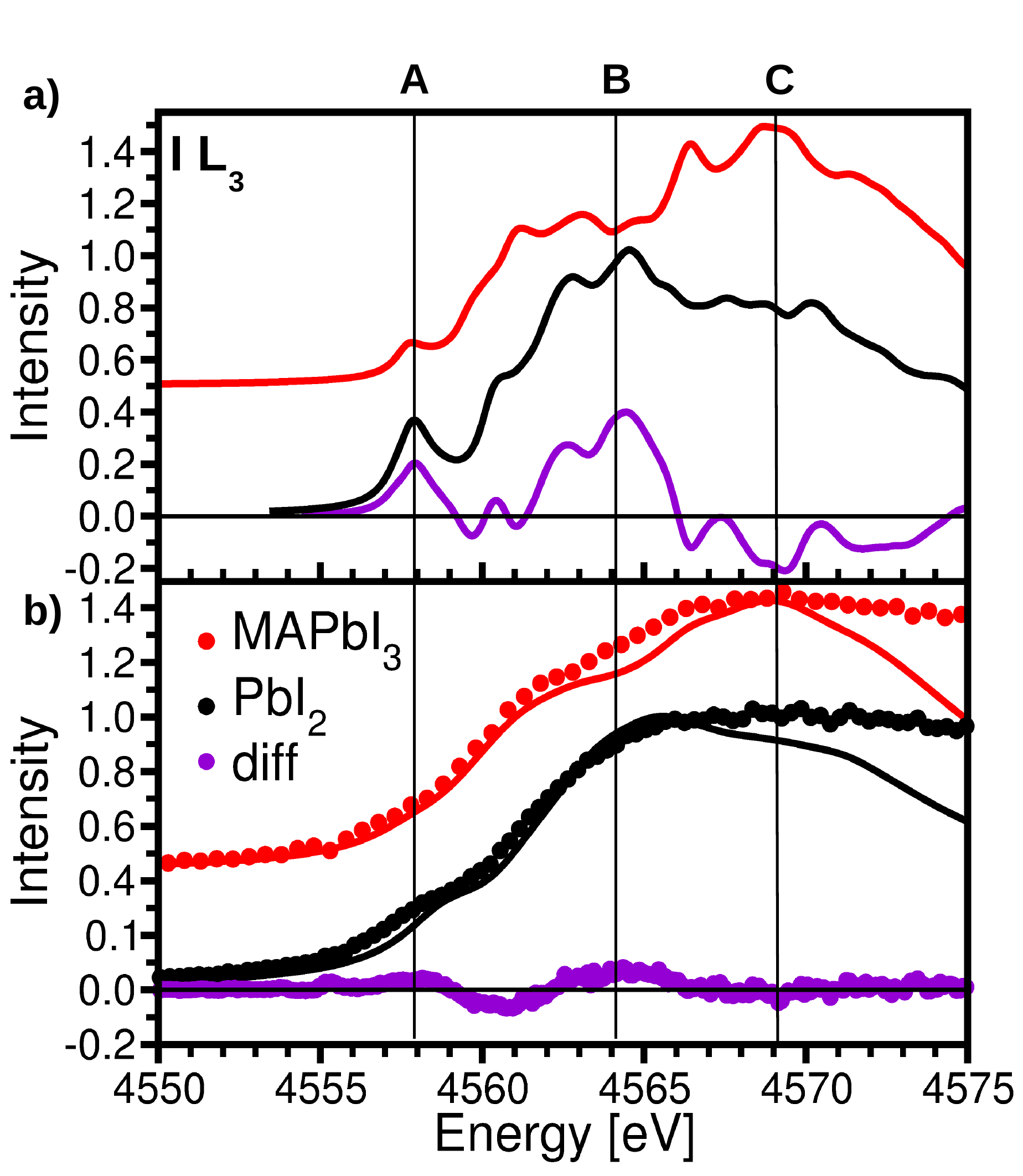}
\caption{\label{fig1:L3} a) Calculated X-ray absorption near-edge spectra of the I $L_3$ edge of \ce{MAPbI3} (red) and \ce{PbI2} (black). A Lorentzian broadening of 0.5 eV is used in the calculations. b) Experimental XANES from the $L_3$ egde. For a direct comparison, calculated spectra with an increased broadening of 1.5 eV are included. The spectra of \ce{MAPbI3} are offset by 0.5 for better readability. In both panels the difference between the spectra of \ce{PbI2} and \ce{MAPbI3} is shown in purple. The normalization of the calculated and experimental spectra is described in the Supporting Information.
}
\end{figure}

We start our analysis by examining the XANES of \ce{MAPbI3} and \ce{PbI2} from the I $L_3$ edge, shown in Fig.~\ref{fig1:L3}.
Due to the large spin-orbit splitting of 298.51 eV between I $2p_{3/2}$ and $2p_{1/2}$ electrons, the $L_2$ and $L_3$ edges can be treated independently within our theoretical framework (see Fig.~S2 in the Supporting Information [SI]). The spectra of both materials (Fig.~\ref{fig1:L3}a) are characterized by the three main features, labeled A, B, and C, with only some differences in the intensity distribution. Our many-body perturbation theory (MBPT) results reveal that the feature A is formed by two bound excitons with binding energies of 480 meV in \ce{MAPbI3} and 450 meV in \ce{PbI2}.
At higher transition energies, where peaks B and C appear, more pronounced differences in terms of oscillator strength are visible in the calculated XANES of the two compounds.
The features A and B stem from interband transitions to the unoccupied I $d$ bands, and thus the intensity of these excitations reflects the differences in the electronic structure of the two materials. Note that in the calculated XANES a relatively small Lorentzian broadening of 0.5 eV is chosen to better resolve the fine structure of the absorption rise and the presence of the two above-mentioned resonances.
A detailed analysis of the element-projected density of states (DOS) of both \ce{MAPbI3} and \ce{PbI2} is reported in the SI (Fig.~S1).
In the experimental spectrum (Fig.~\ref{fig1:L3}b), the fine structure is masked by the considerable intrinsic lifetime broadening, which amounts to at least 3.08 eV at the I $L_3$ edge~\cite{natwidth, atomdata}. For a better comparison with the experimental data in Fig.~\ref{fig1:L3}b, the calculated spectra are plotted with an increased broadening of 1.5 eV.   
Additional information illustrating the good agreement between theory and experiment is obtained by considering the difference between the spectra of \ce{MAPbI3} and \ce{PbI2}, as shown in Fig.~\ref{fig1:L3} in purple color. 
In this way, peaks A and B become visible also in the experimental results (Fig.~\ref{fig1:L3}b). Indeed, the computed spectra (Fig.~\ref{fig1:L3}a) reproduce not only the relative position of the maxima, but also the minimum at about 4561 eV. 
The overall magnitude of the computed signal is larger than in the experimental one due to the small lifetime broadening.
\begin{figure}[t]
\includegraphics[width=0.425\linewidth]{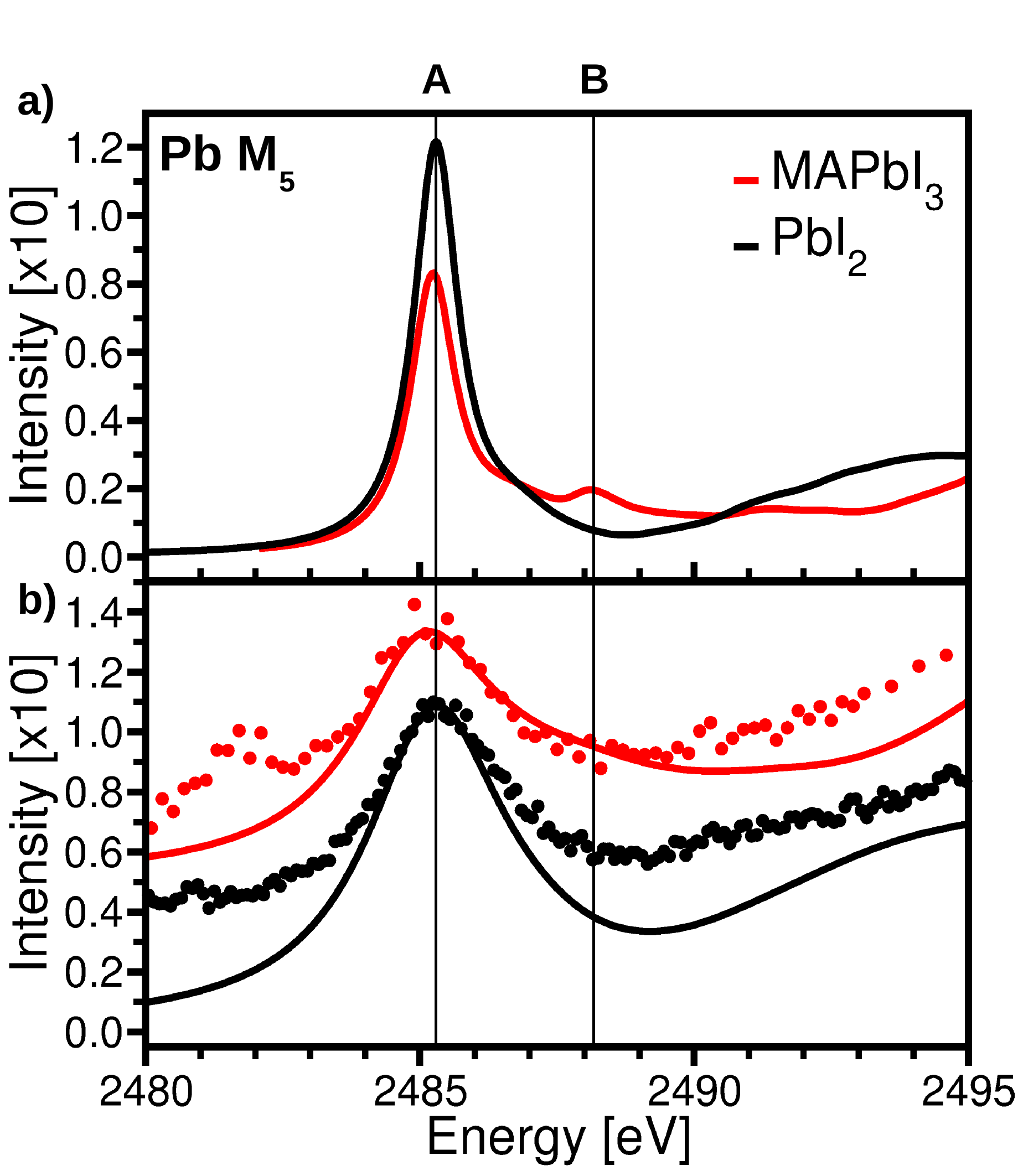}
\caption{\label{fig3:M5} X-ray absorption near-edge spectra of the Pb $M_5$ edge of \ce{MAPbI3} (red) and \ce{PbI2} (black) as obtained from (a) theory and (b) experiment. In panel (a), a Lorentzian broadening of 0.5 eV is included. For a direct comparison, panel (b) contains calculated spectra with an increased Lorentzian broadening of 1.5 eV. The spectra of \ce{MAPbI3} are offset in intensity by 0.5 for better readability.}
\end{figure}
In contrast to the results from the I $L_3$ edge spectra discussed above, the Pb $M_5$ edge spectra of \ce{MAPbI3} and \ce{PbI2} shown in Fig.~\ref{fig3:M5} exhibit a distinct pre-edge feature both in theory (Fig.~\ref{fig3:M5}a) and experiment (Fig.~\ref{fig3:M5}b). This pronounced peak at 2485 eV has a clear excitonic nature. From MBPT calculations we identify  this peak as formed by several bound electron-hole pairs with binding energies up to 860 meV in \ce{MAPbI3} and 950 meV in \ce{PbI2} (see also the discussion below related to Fig.~\ref{fig5:BSEvsIP}). At higher energies, around 2488 eV, we find an additional peak in the calculated spectrum of \ce{MAPbI3}, which does not appear in that of \ce{PbI2}. From the analysis of the projected DOS (see SI, Fig.~S1), we can clarify that this weak peak in the perovskite originates from transitions to the Pb $p$ states above the band gap, which are present only in the electronic structure of \ce{MAPbI3} but absent in the one of \ce{PbI2}. 
In the experimental spectrum of \ce{MAPbI3} (Fig.~\ref{fig3:M5}b) this peak is not visible, most likely due to the large lifetime broadening of about 3 eV~\cite{atomdata}, that characterizes also the excitations from the Pb $3d$ electrons. For direct comparison, Fig.~\ref{fig3:M5}b also displays calculated spectra with an increased Lorentzian broadening of 1.5 eV. As expected, the feature B is masked by the broadening.
The measured XANES of \ce{MAPbI3} displays an additional pre-edge peak at about 2481 eV, which does not appear in the spectrum of \ce{PbI2} and is not reproduced by theory. Additional calculations (details reported in Fig.~S3 in the SI) rule out effects due to a phase transition in the perovskite sample, as well as to the presence of metallic elemental Pb, which can be formed upon X-ray illumination, as reported in the recent literature~\cite{degradation:light, degradation:x-ray, bestpractices}. X-ray transitions ascribed to \ce{Pb}-\ce{SO4} and \ce{Pb}-\ce{O} bonds are reported to be in this spectral range \cite{XANESLead}, however, hard X-ray photoemission data (not shown) does not show any evidence for the presence of either sulfur and/or \ce{Pb}-\ce{O} at the surface of the studied samples.
Therefore, some uncertainty remains on the origin of the pre-peak in the Pb $M_5$ edge spectrum of \ce{MAPbI3}.

The analysis of the XANES presented above shows remarkable similarities between the spectra of \ce{MAPbI3} and \ce{PbI2} for both considered absorption edges.
While, on the one hand, this result may discourage the use of X-ray absorption techniques to detect the presence of residual \ce{PbI2} in the hybrid perovskite samples, on the other hand, it confirms that the spectral fingerprints of \ce{MAPbI3} are to a large extent determined by its inorganic cage. Our \textit{ab initio} many-body approach can be exploited to further characterize the electron-hole pairs that dominate the absorption onset of the XANES of \ce{MAPbI3} and \ce{PbI2} from both I $L_3$ and Pb $M_5$ edges.

In Fig.~\ref{fig5:BSEvsIP}, we report the spectra of \ce{MAPbI3} computed with and without the inclusion of electron-hole correlation, namely by solving the Bethe-Salpeter equation (BSE) and within the independent-particle approximation (IPA). In both panels, the dashed line indicates the onset of the IPA spectrum. For the bound excitons in the BSE spectra, the binding energy $E_b$ is obtained as the difference with respect to the IPA onset. As already mentioned above, we find $E_b=480$ meV and $E_b=860$ meV for the lowest-energy exciton in the I $L_3$ and Pb $M_5$ edge XANES of \ce{MAPbI3}, respectively. The binding energies of the lowest-energy bound exciton in the Pb $M_5$ edge of \ce{PbI2}  is very similar to the one in \ce{MAPbI3}, as reported recently~\cite{Vorwerk2017}. Excitons in core spectra are considerably more strongly bound than those in the optical spectra, resulting in significantly larger binding energies compared to those for of valence excitons, for which binding energies of 40~meV have been reported for \ce{MAPbI3} based on BSE calculations \cite{Zu2014}. This binding energy difference is mainly due to the localization of the involved core states.
\begin{figure}[t]
\includegraphics[width=0.425\linewidth]{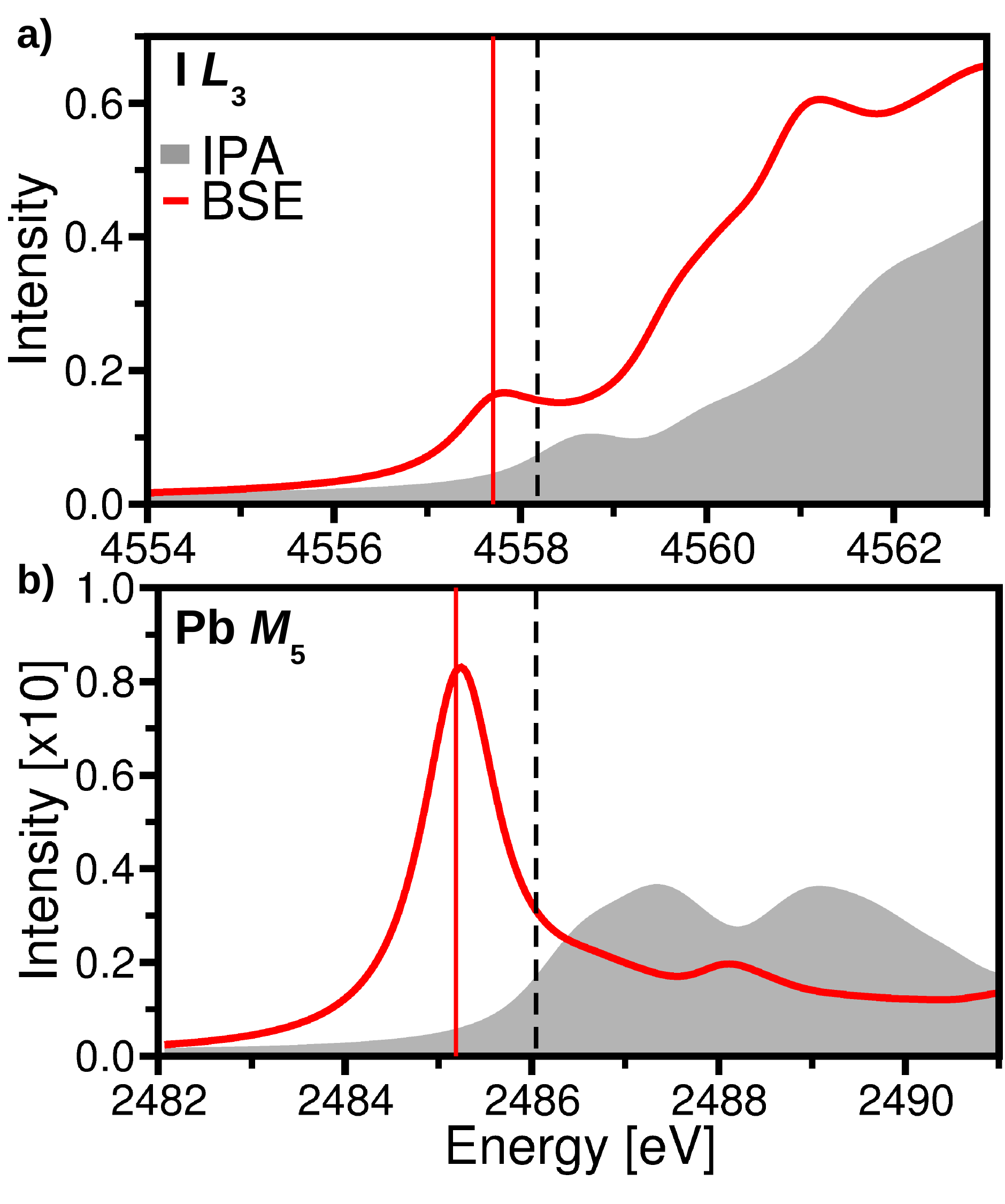}
\caption{\label{fig5:BSEvsIP}X-ray absorption near-edge spectrum from the a) I $L_3$ and b) Pb $M_5$ edge of \ce{MAPbI3} as obtained from the solution of the Bethe-Salpeter equation (BSE, red line) and within the independent-particle approximation (IPA, shaded area). The onset of the IPA spectrum is marked by the vertical dashed line, while the position of the lowest-energy bound exciton from BSE by the solid red line.}
\end{figure}
%

%
\begin{figure}[t]
\includegraphics[width=0.98\linewidth]{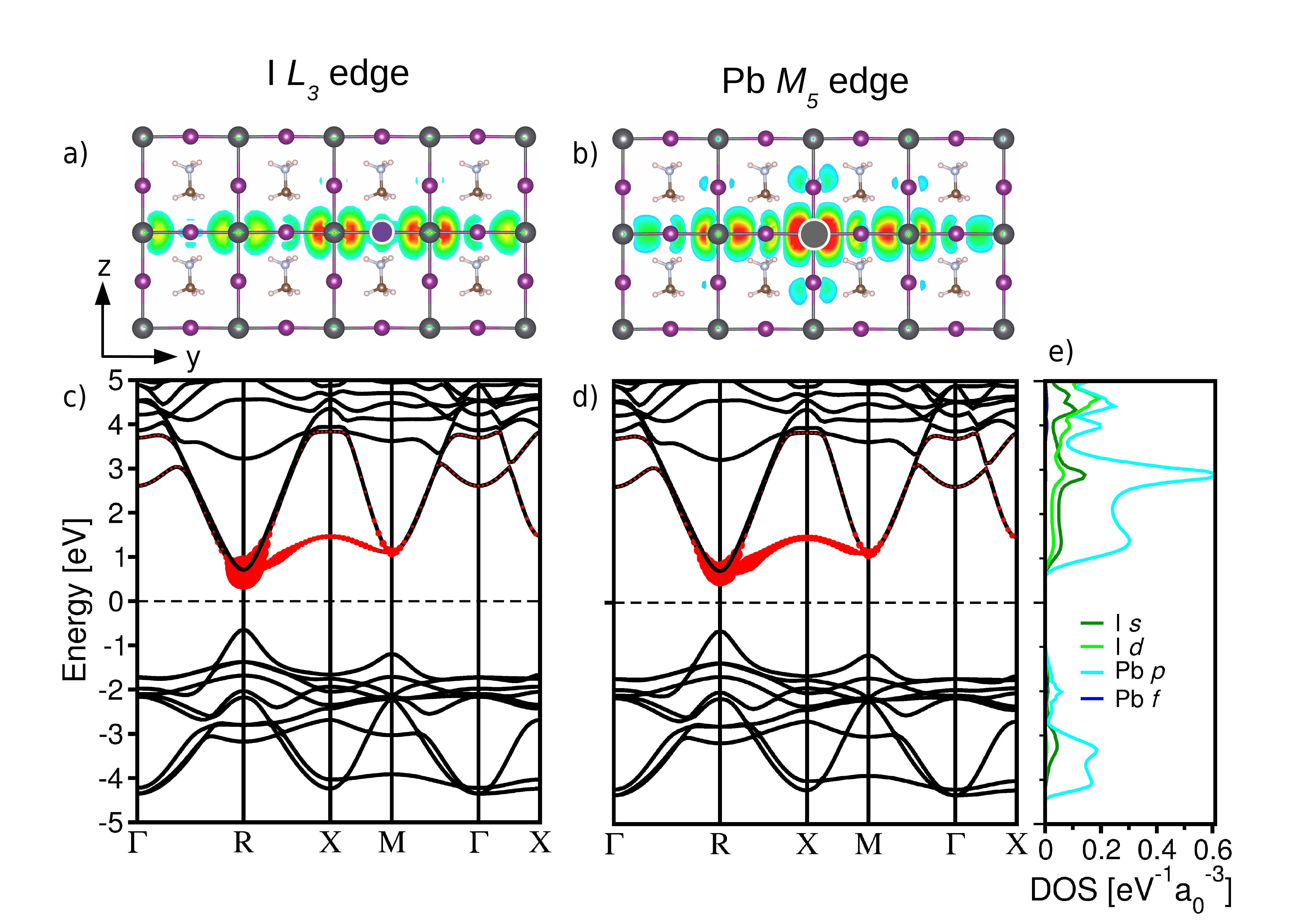}
\caption{\label{fig4:M_contr} Real-space distribution of the first exciton of \ce{MAPbI3} in the XANES from a) the I~$L_3$ and b) the Pb~$M_5$ edge. The position of the core hole is fixed at the I and Pb atoms, respectively, and marked in both cases by a large dot in the color of the corresponding atom (purple for I and grey for Pb). 
c) Reciprocal-space analysis the first exciton of \ce{MAPbI3} from c) the I~$L_3$ and d) the Pb~$M_5$ edge. The size of the red circles quantifies the contribution of each electronic state to the exciton. The Fermi energy is set to zero in the mid-gap and marked by a dashed line. e) Projected density of states of \ce{MAPbI3}, with relevant contributions from the I $s$ and $d$ states, as well as Pb $p$ and $f$ states.}
\end{figure}

In Fig.~\ref{fig4:M_contr}, we show the real- and reciprocal-space representation of the lowest-energy excitons in the XANES of \ce{MAPbI3} from the I $L_3$ and the Pb $M_5$ edge. The excitonic wave-functions displayed in Fig.~\ref{fig4:M_contr}a,b are remarkably similar.
With the hole fixed at the position of the I and Pb atom, respectively (more details in the SI), the electron distribution is extended along the Pb-I bond ($y$ axis, see Fig.~\ref{fig4:M_contr}a) over several unit cells, while in the directions perpendicular to this bond it is localized within one unit cell. 
This anisotropy is not visible in the absorption spectra, because the excitons extending along the bond are almost degenerate.
The probability density associated with the excitonic wave-function in Fig.~\ref{fig4:M_contr}a exhibits depletion along the Pb-I bond, due to the character of the hybridized Pb-I anti-bonding state formed at the bottom of the conduction band. 
While the wave-function distribution around the absorbing I atom is nearly spherically symmetric, the one around the neighboring Pb and I atoms reveals the contribution of the anisiotropic $p$ states. 
A similar character is found also for the first bound exciton in the spectrum of \ce{MAPbI3} from the Pb $M_5$-edge (Fig.~\ref{fig4:M_contr}b). 
Since the Pb atoms in the center of the octahedron are bound to six I atoms along the three Cartesian directions, a number of quasi-degenerate electron-hole pairs appear. 
Excitons with the largest binding energies, \textit{i.e.}, with the lowest excitation energy, are two-fold degenerate. These two degenerate excitons are therefore symmetric along the Pb-I bonds perpendicular to the molecular axis. In Fig.~\ref{fig4:M_contr}b we show the real-space distribution of the exciton extending within the $(\bar{1}00)$ plane.
Bound excitons generated by radiation with polarization parallel to it have slightly smaller binding energies. 
Again, the anisotropic anti-bonding character directly stemming from the targeted unoccupied electronic states is clearly visible. The distribution of this electron-hole pair is also very similar to the one obtained for the Pb $M_{4,5}$-edge spectrum of \ce{PbI2}~\cite{Vorwerk2017}. In that case, an analogous distribution along the Pb-I bonding direction is shown. However, a higher degree of in-plane localization appears in \ce{PbI2}, where the exciton is confined within a single layer.

The real-space distribution of the excitons is obviously reflected also in reciprocal space, as shown by the band-structure plots in Fig.~\ref{fig4:M_contr}c,d. 
The contributions to both excitons arise mostly from the bottom of the conduction band at the R point, where the direct band gap of \ce{MAPbI3} appears. Much smaller contributions originate from the R-X path. This distribution of core excitons in reciprocal space agrees very well with the one shown for bound excitons in the optical spectrum of \ce{MAPbI3} obtained on the same level of theory \cite{Kresse2016}.
The two lowest unoccupied bands, where most of the excitonic weight is localized, are characterized by a strong hybridization between Pb and I $p$ states, as shown in the projected DOS in Fig.~\ref{fig4:M_contr}e. 
This hybridization is reflected in the difference in peak intensity between the two excitons: Transitions from the Pb $3d_{5/2}$ ($M_5$ edge) spectrum target directly the Pb $p$ contributions to the conduction band, giving rise to the intense peak in the spectrum. 
On the other hand, transitions from the I $2p_{3/2}$ levels ($L_3$ spectrum) are dipole-allowed only for the unoccupied I states with $s$ and $d$ character, which contribute to a much smaller extent to the lowest conduction band (Fig.~\ref{fig4:M_contr}e). 
This explains the relatively low oscillator strength of the first peak in the I $L_3$ spectrum of \ce{MAPbI3} (Fig.~\ref{fig1:L3}a) compared to the one at the onset of the $M_5$ edge (Fig.~\ref{fig3:M5}a).

To summarize, in a joint theoretical and experimental work we have studied core level excitations from the I $L_3$ and the Pb $M_5$ edge of \ce{MAPbI3} and its binary phase \ce{PbI2}. We have shown that these two materials exhibit very similar XANES at both edges, suggesting the dominant contribution from the inorganic cage to the core excitations of \ce{MAPbI3}. 
Differences between the spectra concern mainly minor features, which can hardly be exploited to detect the presence of the binary phase in hybrid perovskite samples. 
The absorption onset in the XANES of \ce{MAPbI3} from both considered edges is characterized by pronounced excitonic effects.
A detailed theoretical analysis shows that the bound excitons in the core spectra extend along the Pb-I bond for several unit cells, while being confined in the other directions. They originate from excitations targeting the lowest-energy conduction bands around high-symmetry points in the Brillouin zone. Moreover, the intensity of the peaks reveals the hybridization of iodine and lead derived states. Overall, this combined experimental and theoretical analysis of the core excitations offers a new perspective on the electronic structure  in this complex material and its excitations.
\section{Experimental Methods}
\textit{Theoretical Methods} -- Core-level absorption spectra are obtained in a two-step process: First, the electronic structure is calculated from density-functional theory (DFT)~\cite{hohe-kohn64pr,kohn-sham65pr} employing the generalized gradient approximation for the exchange-correlation functional, as developed by Perdew, Burke and Ernzerhof~\cite{perd+96prl}. 
Second, \textit{ab initio} X-ray spectra are obtained in the framework of many-body perturbation theory (MBPT)\cite{Vorwerk2017} from the solution of the Bethe-Salpeter equation (BSE)~\cite{hank-sham80prb,Strinati}. 
In this approach, the many body problem is mapped into an effective two-particle Hamiltonian entering the eigenvalue equation $\hat{H}^{BSE}_{cu \mathbf{k}, c'u' \mathbf{k}'}A^{\lambda}_{c' u' \mathbf{k}'}=E^{\lambda}A^{\lambda}_{c u \mathbf{k}}$, where $(cu \mathbf{k})$ denotes any transition from a core ($c$) to an unoccupied state ($u$) at $\mathbf{k}$. 
The electron-hole Hamiltonian $\hat{H}^{BSE}=\hat{H}^{diag}+\hat{H}^{x}-\hat{H}^{d}$ accounts for the diagonal term $\hat{H}^{diag}$, which describes single-particle transitions in the independent-particle approximation, the exchange term $\hat{H}^{x}$, which describes the repulsive exchange interaction through the short-range Coulomb potential, and the direct term $H^{d}$, which includes the attractive screened Coulomb interaction between the electron and the hole. Scissors operators are applied to include the quasi-particle correction to the Kohn-Sham states from DFT. Their values of 86 eV for the I $L_3$ and 52 eV for the Pb $M_5$ edge are chosen to align the computed spectra to the experimental ones.
Additional details about the BSE and its implementation within an all-electron full-potential framework are reported in Refs.~\citenum{pusc-ambr02prb,Sagmeister2009}.
The eigenstates $A^{\lambda}$ provide information about the excitonic wavefunction, both in real space, $\Phi^{\lambda}(\mathbf{r}_e,\mathbf{r}_h)=\sum_{c u \mathbf{k}}A^{\lambda}_{cu \mathbf{k}}\psi_{u \mathbf{k}}(\mathbf{r}_e)\psi_{c \mathbf{k}}(\mathbf{r}_h)$, and in reciprocal space through the \textit{exciton weights}, defined as $w^{\lambda}_{u \mathbf{k}}=\sum_{c}|A^{\lambda}_{cu \mathbf{k}}|^2$.
Furthermore, the transition coefficients are obtained as $t^{\lambda}_{i}=\sum_{cu\mathbf{k}}A^{\lambda}_{cu \mathbf{k}}\frac{\langle c\mathbf{k}|p_i| c\mathbf{k} \rangle}{\epsilon_{u \mathbf{k}}-\epsilon_c}$, where $\epsilon_{u \mathbf{k}}$ and $\epsilon_{c}$ are the Kohn-Sham eigenvalues of the conduction and core state respectively. The binding energy of bound excitons is obtained as the difference between the transition energy $E^{\lambda}$ and the onset of independent-particle transitions.
The macroscopic dielectric function $\epsilon_M$ is calculated as $\epsilon_M^{ij}(\omega)=\delta_{ij}+\sum_{\lambda}\frac{t^{\lambda}_{i}\left[t^{\lambda}_j\right]^{*}}{\omega-E^{\lambda}+\textrm{i}\Gamma}$. 
All calculations are performed with the all-electron, full-potential code \texttt{exciting}~\cite{exciting} implementing DFT and MBPT and including an explicit treatment of core electrons. This is done by adopting the linearized augmented plane-wave (LAPW) basis set. Computational details are reported in the SI.

\textit{Experimental Methods} -- \ce{MAPbI3} perovskite thin films of 300 nm nominal thickness were prepared on compact \ce{TiO2}/FTO/glass substrates at University of Oxford following the standard "one-pot" preparation approach \cite{solar,lowtemp}. The compact \ce{TiO2} layers were prepared by spin-coating an acidic solution of titanium isopropoxide dissolved in ethanol at 2000 rpm for 60s on fluorine-doped tin oxide (FTO) substrates (Pilkington, TEC7) followed by drying at $150\; ^{\circ}\textrm{C}$ and annealing at $500 \; ^{\circ}\textrm{C}$ for 45 min. The precursor solution for the perovskite was produced by dissolving methylammonium iodide (\ce{CH3NH3I}, "MAI") and lead (II) chloride (\ce{PbCl2}) in anhydrous N,N-dimethylformamide (DMF) in a 3:1 molar ratio with a final concentration of 2.64 mol/l MAI and 0.88 mol/l \ce{PbCl2}. This solution was spin-coated onto compact \ce{TiO2} at 2000 rpm in a nitrogen-filled glovebox for 45 s. After spin-coating, the films were left to dry at room temperature inside the glovebox to allow the solvent to slowly evaporate, followed by an annealing step for 2.5 h at $90\;^{\circ}\textrm{C}$  (this step is needed for the crystallization and formation of the perovskite structure). After preparation, samples were sealed in a container under inert gas and transferred from the University of Oxford to the Helmholtz Zentrum Berlin f\"ur Materialien und Energie GmbH (HZB), where they were again unpacked and mounted on sample holders in a \ce{N2}-purged glovebox. The samples were introduced into the load lock of the High Kinetic Energy Photoelectron Spectrometer (HiKE) endstation \cite{Gorgoi200948} (see below) with a N$_2$-filled glovebag to minimize exposure to ambient air.
As \ce{PbI2} reference sample commercially available powder (Sigma-Aldrich, 99.999\% trace metals basis) was used.  The powder was mounted in air onto the sample holder by pressing it onto double-sided carbon tape. 

XANES measurements of the I L$_{2,3}$ and Pb $M_{4,5}$ edges were carried out in the HiKE endstation \cite{Gorgoi200948} located at the BESSY II KMC-1 beamline \cite{Schaefers2007} at Helmholtz-Zentrum Berlin (HZB). For the I $L_{2,3}$ edge XANES spectrum the excitation energy was scanned through the range of 4500 -- 4950 eV and for Pb $M_{4,5}$ -- edge through 2400 -- 2650 eV recording the I $\textrm{L}_{\alpha+\beta}$ and  the Pb $\textrm{M}_{\alpha,\beta}$ emissions in partial fluorescence yield (PFY) mode, respectively. The (selected) fluorescence photons were detected with a Bruker XFlash\texttrademark  4010 silicon drift detector with a beryllium window. The energy steps for the scanning of the excitation energy were varied for different energy ranges. For the I $L_3$ edge in Fig.~\ref{fig1:L3}, the data points were measured with an energy step of 0.1 eV for \ce{PbI2} and 0.5 eV for \ce{MAPbI3} in the shown energy range.  For the Pb $M_5$ edge in Fig.~\ref{fig3:M5}, the energy step in the shown excitation energy range was 0.2 eV for \ce{MAPbI3} and 0.1 eV for \ce{PbI2}. For the energy calibration of the photon energy Au $4f$ peaks were always measured on a clean, electrically grounded Au foil, using the starting and ending excitation energies of the I $L_{2,3}$ and Pb $M_{4,5}$ - edge measurements. The Au $4f$ peaks were fitted with the fitting program "Fityk" \cite{fityk} version 0.9.8 using as an approximation a linear background and Voigt functions to fit the spin-orbit split doublet, by fixing the area ratio according to the multiplicy (2j+1) and coupling the peak shape. The spin-orbit splitting for Au $4f$ was set fixed to 3.67 eV \cite{handbook}. The photon energies were then calibrated by setting the Au $4f_{7/2}$ binding energy to 84.00 eV.

\begin{acknowledgement}
Work partly funded by the German Research Foundation (DFG) through the Collaborative Research Center 951 HIOS and the GraFOx Leibniz ScienceCampus. Funding from the Helmholtz Energieallianz is also appreciated. C.V. acknowledges financial support from the Humboldt Research Track Scholarship of the Humboldt Universit\"at zu Berlin. C.H. acknowledges support from the Potsdam University Helmholtz Zentrum Berlin graduate school HyPerCells: Perovskites Basic Research for High Efficiency Solar Cells. C.C. acknowledges support from the Berliner Chancengleichheitsprogramm and IRIS Adlershof. C.H., R.F., R.G.W., and M.B. additionally acknowledge funding from the Helmholtz Association (VH-NG- 423).
We thank Helmholtz Zentrum Berlin for the allocation of synchrotron radiation beamtime for XANES measurements. We thank Karsten Hannewald for fruitful discussions in the early stage of the project.
\end{acknowledgement}

%

\begin{suppinfo}

The Supporting Information contains additional details regarding the structural and electronic properties of \ce{MAPbI3} and \ce{PbI2}, as well as further information about the theoretical spectra.
Details of the first-principles calculations are also reported and Refs.~\citenum{MAPbI3structDFT,PbI2:structure,Pb:structure,kawa+02jpsj} are cited.

\end{suppinfo}




\section*{Supplementary material}

# SUPPORTING INFORMATION

# Exciton-Dominated Core-Level Absorption Spectra of Hybrid Organic-Inorganic Lead Halide Perovskites

Christian Vorwerk,[*,†,‡] Claudia Hartmann,[¶] Caterina Cocchi,[†,‡] Golnaz Sadoughi,[§] Severin N. Habisreutinger,[§,⊥] Roberto Félix,[¶] Regan G. Wilks,[¶,∥] Henry J. Snaith,[§] Marcus Bär,[¶,∥] and Claudia Draxl[†,‡]

†Institut für Physik and IRIS Adlershof, Humboldt-Universität zu Berlin, 12489 Berlin, Germany

‡European Theoretical Spectroscopy Facility

¶Renewable Energy, Helmholtz-Zentrum Berlin für Materialien und Energie GmbH, 14109 Berlin, Germany

§Clarendon Laboratory, Department of Physics, University of Oxford, Oxford OX1 3PU, United Kingdom

∥Energy Materials In-Situ Laboratory Berlin (EMIL), Helmholtz-Zentrum Berlin für Materialien und Energie GmbH, 12489 Berlin, Germany

⊥Chemistry and Nanoscience Center, National Renewable Energy Laboratory (NREL), Golden, CO, USA

E-mail: vorwerk@physik.hu-berlin.de

# Structural Properties of MAPbI$_3$ and PbI$_2$

In this work, we consider the high-temperature cubic phase (space group $Pm\bar{3}m$) of methylammonium lead iodide (MAPbI$_3$). Lattice parameter $a = 6.3115$ Å and atomic coordinates from experiment[1] are used. Since the positions of the hydrogen atoms cannot be resolved in X-ray-based crystallographic methods,[2] these species are manually in the unit cells added to form bonds with C and N atoms. The geometry of the organic group is relaxed until the residual forces are smaller than 0.02 eV/Å, with the coordinates of all the other atoms in the unit cell held fixed. Since the orientation of the organic group is known to have only a minor influence on the electronic properties of the cubic phase of MAPbI$_3$,[3] different molecular orientations are not further explored in this work.

To address the effects of the crystal structure on the core excitations of MAPbI$_3$, in addition to the cubic structure we also study the *tetragonal* phase.[4] In particular, we consider the *reduced* tetragonal phase, which is obtained by including hydrogen atoms in the organic groups. Their presence breaks the *I4cm* symmetry of the tetragonal lattice, thereby reducing the primitive cell from four to two chemical units. In our calculations, we enforce the *I4cm* symmetry before performing a full atomic relaxation. With this assumption, we neglect the influence of different relative molecular orientations that can occur in the full tetragonal phase. This choice is justified, as the influence of molecular orientation on the electronic structure as well as core absorption spectra is negligible.

PbI$_2$ is a layered semiconducting material with hexagonal crystal structure (space group $P\bar{3}m1$).[5] Structure optimization yields lattice parameters $a = 4.54$ Å and $c = 6.98$ Å in good agreement with experiments.[5]

We also consider metallic Pb in its cubic phase (space group $Fm\bar{3}n$) for further analyzing the Pb $M_5$-edge absorption spectrum of MAPbI$_3$. For this purpose, the experimental lattice parameter[6] $a = 2.465$ Å is used.

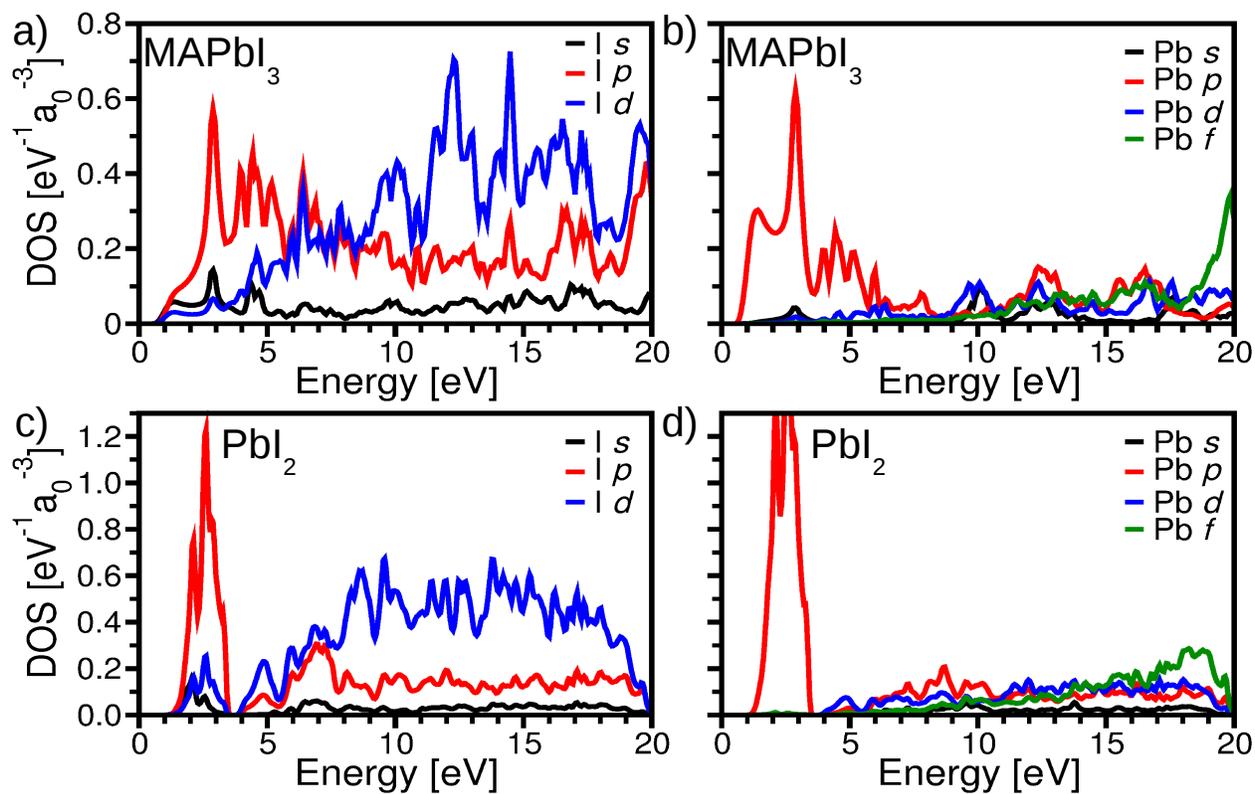

Figure S1: Projected density of states (DOS) of the conduction region of MAPbI$_3$ from a) I and b) Pb states, as well as of PbI$_2$ with contributions from c) I and d) Pb states. The Fermi energy is set to zero in the mid-gap.

# Electronic Structure of MAPbI$_3$ and PbI$_2$

For a better understanding of the core-excitation spectra it is instructive to analyze the projected density of states (PDOS) in the unoccupied region of both MAPbI$_3$ and PbI$_2$ (see Fig. S1). In both materials, the bottom of the conduction band has predominant I and Pb $p$-like character. At about 5 eV above the Fermi energy ($E_F$) the contribution from I $d$ states become larger than the I $p$ states. The PDOS of Pb shows predominately $p$-like character at low energies. In MAPbI$_3$, the Pb-$p$ PDOS has two distinct peaks at approximately 1 eV and 3 eV, respectively (Fig. S1b), while in PbI$_2$ the Pb-$p$ PDOS exhibits only one peak at approximately 3 eV above $E_F$ (Fig. S1d). This difference in the PDOS is reflected in the core excitations from the Pb $M_5$ edge, where a peak appears in the spectrum of MAPbI$_3$, while being absent in the one of PbI$_2$.

# Calculated X-Ray Absorption Spectra of MAPbI$_3$ and PbI$_2$

In Fig. S2 we report the X-ray absorption spectra from the I $L_{2,3}$ and Pb $M_{4,5}$ edges for both MAPbI$_3$ and PbI$_2$. Due to the large spin-orbit splitting of 298.51 eV between the I $2p_{3/2}$ and $2p_{1/2}$ states and 104 eV between the Pb $3d_{3/2}$ and Pb $3d_{5/2}$ states, transition from the I $L_2$ and $L_3$ as well as from the Pb $M_4$ and Pb $M_5$ edges can be treated separately in our theoretical framework. Spectra are computed considering only one absorbing atom, since different atomic contributions yield the same result in both materials. Only the $xx$-component of the dielectric tensor is displayed, as anisotropic effects are masked by the broadening. The plots report *raw theoretical spectra*, which are therefore not aligned in energy nor normalized in intensity compared to the experimental values. Thus, the scale of the y-axis differs between Fig. S2 and Fig. 1. It is evident from the results shown in Fig. S2 that the spectral features in the I $L_2$ and $L_3$ as well as in the Pb $M_4$ and $M_5$ sub-edges are identical. Excitation energies are obviously different, due to the different initial states of the excited core electrons.

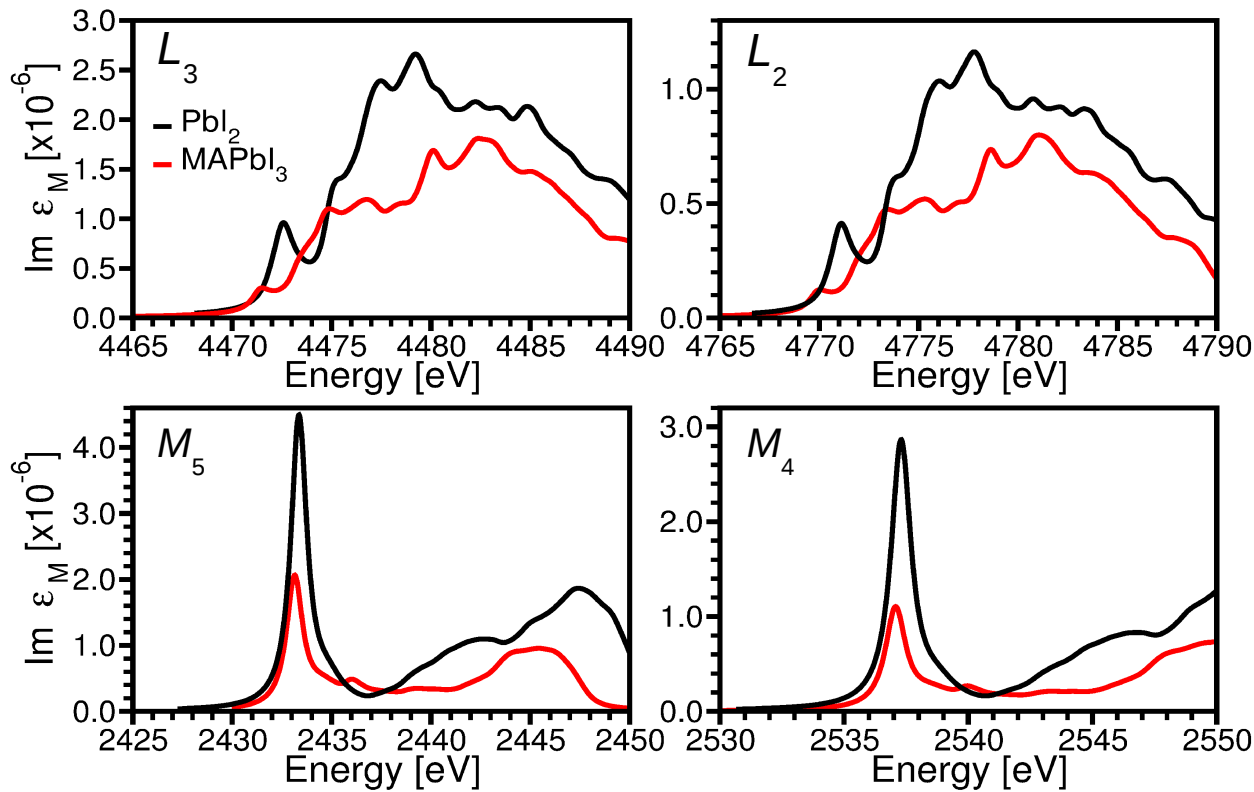

Figure S2: X-ray absorption near-edge spectra of MAPbI$_3$ (red) and PbI$_2$ (black) from the absorption edges indicated in the respective panel. For the I $L_2$ and $L_3$ edges, the spectrum of PbI$_2$ is scaled by a factor 2/3, to account for the different number of iodide atoms per unit cell compared to MAPbI$_3$.

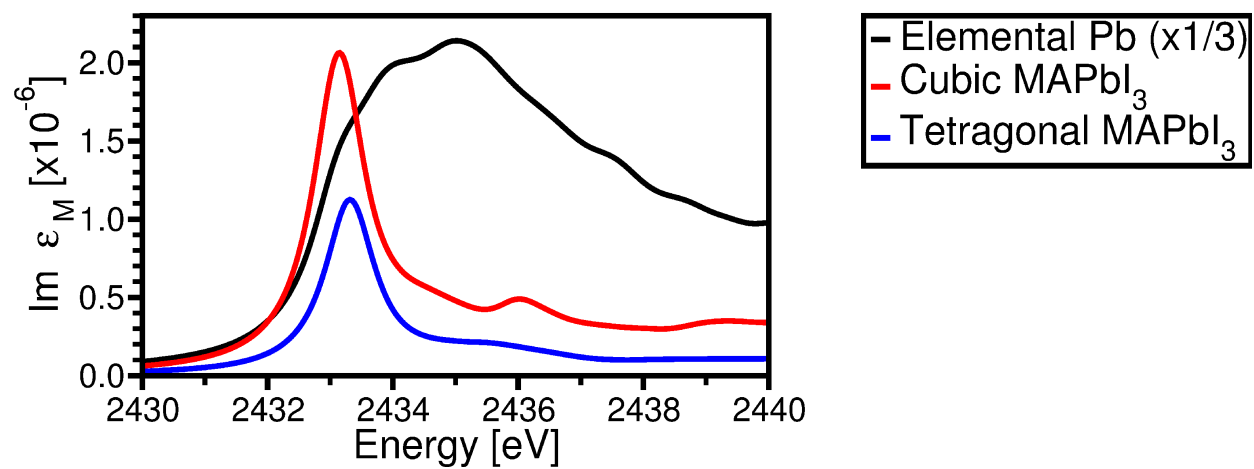

Figure S3: Pb $M_5$ edge XANES of cubic (red) and tetragonal (blue) MAPbI$_3$ and of elemental lead (black). The spectrum of elemental lead is reduced by a factor of 3 for better comparison with the others.

In Fig. S3, the Pb $M_5$ edge XANES of the tetragonal and cubic phase of MAPbI$_3$ and the corresponding XANES of elemental lead are displayed. As in Fig. S2, we report *raw theoretical spectra*, which are neither aligned nor renormalized to the experimental ones. The peak at approximately 2433 eV, which is formed by several bound excitons, is visible in the spectrum of both phases of MAPbI$_3$. The real-space distribution of the exciton in the cubic phase is shown in Fig. 4b). A similar distribution is also expected for the exciton in the tetragonal phase. Remarkably, the intense peak is missing in the spectrum of elemental lead. As elemental lead is metallic, the electron-hole interaction is largely reduced by the strong screening, and bound excitons are not observed. Note that the pre-peak feature observed in experiment (see Fig. 2b in the main text) is not observed in any of the spectra reported in Fig. S3. These results rule out that the binary phase PbI$_2$ or elemental Pb are responsible for this feature.

## Alignment of Calculated XANES to Experimental Spectra

Scissors operators are applied to the calculated x-ray spectra to account for quasi-particle corrections of both core and conduction single-particle energies. The value of the adopted scissors operator is mainly given by the rigid shift of the core states and is chosen such that a selected feature of the calculated spectra is aligned with the corresponding feature in the experimental one. A value of 52.11 eV is used for the Pb $M_5$ edge spectrum of both MAPbI$_3$ and PbI$_2$, and a value of 86.32 eV for the I $L_3$ edge spectra of both materials. An identical scissors operator can be adopted for both systems due to the considerably smaller core-level shifts of 780 meV between the Pb $3d_{5/2}$ and 470 meV between the I $2p_{3/2}$ states of the two systems.

## Computational Details

Ground-state calculations of the cubic phase of MAPbI$_3$ are performed with a $4 \times 4 \times 4$ **k**-grid and a plane-wave cutoff $R_{MT,max}|\mathbf{G}+\mathbf{q}|_{max} = 4.7$. Muffin-tin (MT) spheres of radius $R_{MT}(Pb) = R_{MT}(I) = 2.2\ a_0$, $R_{MT}(C) = R_{MT}(N) = 1.0\ a_0$, and $R_{MT}(H) = 0.8\ a_0$ are employed.

For the calculation of the spectrum from the I $L_3$ edge, a $4 \times 4 \times 4$ **q**-grid shifted by $\Delta \mathbf{q} = (0.05, 0.15, 0.25)$ and $R_{MT,max}|\mathbf{G}+\mathbf{q}|_{max} = 4.7$ are used. Local field effects (LFE) are included in the BSE calculations up to a cut-off $|\mathbf{G}+\mathbf{q}|_{max} = 2.0$. The screening of the Coulomb interaction is calculated in the random-phase approximation (RPA), including all valence and 100 unoccupied bands. 70 unoccupied bands are included in the diagonalization of the BSE Hamiltonian. The calculated spectrum shown in the main text is shifted by 86.23 eV to align the most intense calculated peak to the most intense one in the experimental spectrum. The calculated spectrum is normalized to the intensity of this most intense peak. For the Pb $M_5$ edge, a $6 \times 6 \times 6$ **q**-grid shifted by $\Delta \mathbf{q} = (0.05, 0.15, 0.25)$ and $R_{MT,max}|\mathbf{G}+\mathbf{q}|_{max} = 4.7$ are adopted. The RPA screening is obtained as the one for the I $L_3$ edge. 40 unoccupied bands are included in the diagonalization of the BSE Hamiltonian. The calculated spectrum in the main text is shifted by 52.11 eV in order to align the most intense peak of the calculated spectrum to the one in the experimental spectrum. As for the previous edge, the spectrum is normalized to the intensity of the most pronounced peak in the experimental spectrum. BSE calculations for all x-ray absorption spectra are converged with respect to the parameters listed below. The real-space distribution of the electronic part of the excitonic wavefunction is calculated and plotted for the first (eighth) BSE solution of the I $L_3$ (Pb $M_5$) edge spectrum. In both cases the coordinate of the core-hole is fixed slightly off with respect to the the position of the absorbing Pb atom, in order to avoid the node of the electronic wave-function. In Fig. 4c) and d), the exciton weights $w_{u\mathbf{k}}^{\lambda} = \sum_c |A_{cu\mathbf{k}}^{\lambda}|^2$ are interpolated on 100 **k**-points along the high-symmetry path in reciprocal space.

For the reduced tetragonal phase of MAPbI$_3$, ground-state calculations are performed on

a $4\times4\times4$ **k**-grid using plane-wave cutoff $R_{MT,max}|\mathbf{G}+\mathbf{q}|_{max} = 4.4$. The MT radius of C and N is 1.1 $a_0$, while all other radii are equivalent to those used in the cubic phase. The spectrum from the Pb $M_5$ edge is obtained on a $4\times4\times4$ **q**-grid shifted by $\Delta\mathbf{q} = (0.05, 0.15, 0.25)$, with $R_{MT,max}|\mathbf{G}+\mathbf{q}|_{max} = 4$. LFE are accounted for up to a cut-off of $|\mathbf{G}+\mathbf{q}|_{max} = 2.5$ $a_0^{-1}$. The RPA screening is computed with all valence bands and 75 unoccupied ones. 40 unoccupied bands are included in the diagonalization of the BSE Hamiltonian.

For PbI$_2$, ground-state calculations are performed with a **k**-grid of $5\times5\times3$ and $R_{MT,max}|\mathbf{G}+\mathbf{q}|_{max} = 10$. MT spheres of $R_{MT} = 2.6$ $a_0$ are used for both species. For the calculation of the XANES from the I $L_3$ edge, a $7 \times 7 \times 5$ **q**-grid shifted by $\Delta\mathbf{q} = (0.05, 0.15, 0.25)$, $R_{MT,max}|\mathbf{G}+\mathbf{q}|_{max} = 11.5$, and $|\mathbf{G}+\mathbf{q}|_{max} = 3.0$ $a_0^{-1}$ are used. 100 empty bands are included in the calculation of the RPA screening and the BSE Hamiltonian includes 40 unoccupied bands. The XANES from the Pb $M_5$ edge is calculated on a $8 \times 8 \times 6$ **q**-grid shifted by $\Delta\mathbf{q} = (0.05, 0.15, 0.25)$, with $R_{MT,max}|\mathbf{G}+\mathbf{q}|_{max} = 10.0$ and $|\mathbf{G}+\mathbf{q}|_{max} = 2.5$ $a_0^{-1}$. The same RPA screening as for the XANES from the I $L_3$ edge is used. The BSE Hamiltonian includes 25 unoccupied bands. The calculated spectra for the Pb $M_5$ and I $L_3$ are shifted by 52.11 eV and 86.32 eV, respectively, such that the most intense peak in the calculated spectrum is aligned with the most intense one in the experimental spectrum. As for MAPbI$_3$, the calculated spectrum is normalized to the height of the most intense peak.

The ground-state properties of elemental Pb are determined on a $7 \times 7 \times 7$ **k**-grid with a plane-wave cut-off $R_{MT,max}|\mathbf{G}+\mathbf{q}|_{max} = 7.0$. For the BSE calculations from the Pb $M_5$ edge, a shifted $12 \times 12 \times 12$ **q**-grid with cut-offs $R_{MT,max}|\mathbf{G}+\mathbf{q}|_{max} = 7.0$ and $|\mathbf{G}+\mathbf{q}|_{max} = 2.0$ $a_0^{-1}$ is used. The RPA screening involves 30 unoccupied states, while 10 unoccupied states are included in the BSE Hamiltonian.

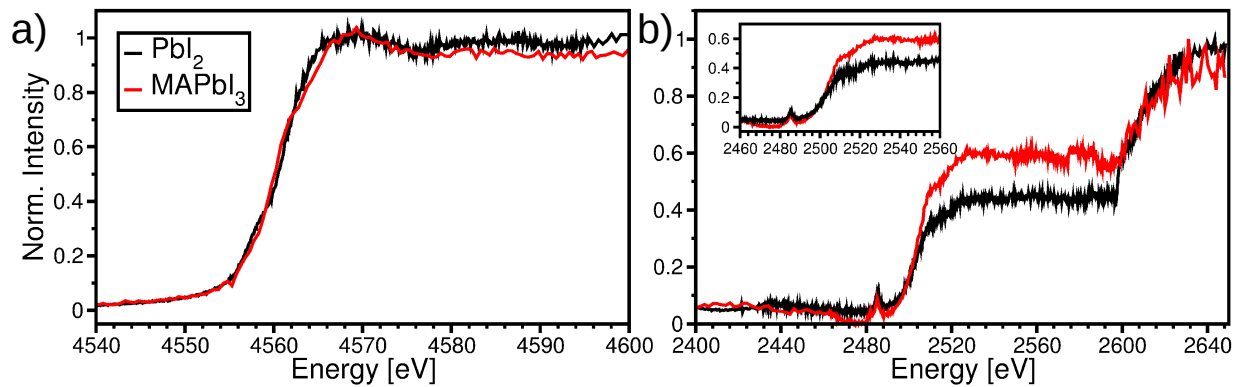

Figure S4: Full measurement window of the experimentally derived a) I $L_3$ and b) Pb $M_{4,5}$ edge XANES spectra of MAPbI$_3$ (red) and PbI$_2$ (black). The inset in b) shows the Pb $M_5$ sub-edge XANES.

# Experimental Data

Figure S4 a) and b) show the XANES I $L_3$ edge and XANES Pb $M_{4,5}$ edge raw data of MAPbI$_3$ and PbI$_2$ reference sample measured in PFY mode. As depicted, the spectra are only normalized to the ionization current (I$_0$) and to the edge jump. As visible from the different signal-to-noise ratio, the spectra (and even different ranges within the spectra – most apparent for the PbI$_2$ data) have been measured with different step sizes and dwell times in order to optimize use of measurement time.

 
\end{document}